\documentclass[aps,pra,reprint,superscriptaddress]{revtex4-2}
\pdfoutput=1
\usepackage{standalone}
\usepackage{nccfoots}
\usepackage{comment}
\usepackage{graphicx}
\usepackage{braket}
\usepackage{bbold}
\usepackage{color}
\usepackage{mathtools}

\definecolor{mygreen}{rgb}{0,0.5,0}
\definecolor{myblue}{rgb}{0,0,0.75}
\definecolor{mymagenta}{cmyk}{0,1,0,0.12}
\usepackage[colorlinks=true,
   linkcolor=blue,
   citecolor=blue,
   urlcolor =blue]{hyperref}

\newcommand{\citeSM}{\cite[{\tiny SM}\kern-0.3em][]{SM}}

\newcommand{\be}{\begin{equation}}
\newcommand{\ee}{\end{equation}}

\newcommand{\tr}{\mathrm{Tr}}
\newcommand{\Var}{\mathbb{V}}
\newcommand{\s}{\mathbf{s}}
\newcommand{\expp}{\mathbb{E}}

\usepackage{amsmath,amsfonts,amssymb,amsthm, bbm,braket}

\expandafter\let\csname equation*\endcsname\relax

\expandafter\let\csname endequation*\endcsname\relax

\usepackage{amsmath}

\usepackage[normalem]{ulem}

\begin{document}

\title{Enhanced estimation of quantum properties with common randomized measurements}
\author{Beno\^it Vermersch}
\email{benoit.vermersch@lpmmc.cnrs.fr}
\affiliation{Univ.\ Grenoble Alpes, CNRS, LPMMC, 38000 Grenoble, France}
\affiliation{Institute for Theoretical Physics, University of Innsbruck, Innsbruck A-6020, Austria}
\affiliation{Institute for Quantum Optics and Quantum Information of the Austrian Academy of Sciences, Innsbruck A-6020, Austria}

\author{Aniket Rath}
\affiliation{Univ.\ Grenoble Alpes, CNRS, LPMMC, 38000 Grenoble, France}

\author{Bharathan Sundar}
\affiliation{Institute for Quantum Information and Matter, Caltech, Pasadena, CA, USA}

\author{Cyril Branciard}
\affiliation{Universit\'e Grenoble Alpes, CNRS, Grenoble INP, Institut N\'eel, 38000 Grenoble, France}

\author{John Preskill}
\affiliation{Institute for Quantum Information and Matter, Caltech, Pasadena, CA, USA}
\affiliation{Walter Burke Institute for Theoretical Physics, Caltech, Pasadena, CA, USA}
\affiliation{Department of Computing and Mathematical Sciences, Caltech, Pasadena, CA, USA}
\affiliation{AWS  Center  for  Quantum  Computing,  Pasadena,  CA, USA}

\author{Andreas Elben}
\email{aelben@caltech.edu}
\affiliation{Institute for Quantum Information and Matter, Caltech, Pasadena, CA, USA}
\affiliation{Walter Burke Institute for Theoretical Physics, Caltech, Pasadena, CA, USA}

\date{\today}

 \begin{abstract}
We present a technique for enhancing the estimation of quantum state properties by incorporating approximate prior knowledge about the quantum state of interest.
This method involves performing randomized measurements on a quantum processor and comparing the results with  those obtained from a classical computer 
that stores an approximation of the quantum state.
We provide unbiased estimators for expectation values of multi-copy observables and present performance guarantees 
in terms of variance bounds which depend on the prior knowledge accuracy.
We demonstrate the effectiveness of our approach through numerical experiments estimating polynomial approximations of the von Neumann entropy and quantum state fidelities.
\end{abstract}

\maketitle

{\it Introduction---}
Classical shadows~\cite{huang2020predicting} have recently emerged as a key element in the randomized measurement (RM) toolbox~\cite{elben2022therandomized}. Previous RM protocols~\cite{vanenk2012measuring,elben2018renyi,elben2020cross,elben2020many} focused on estimating quantum state properties expressible as polynomial functions of a density matrix $\rho$.
Classical shadows enable efficient access to the expectation values $\tr(O\rho)$ of few-body observables $O$. This is particularly important in the context of the variational quantum eigensolver algorithm, which typically requires the measurement of a local Hamiltonian~\cite{Peruzzo2014,cerezo_variational_2021}. More generally,  classical shadows provide access to multi-copy observables (MCO) $\tr(O\rho^{\otimes n})$ ($n\ge 1$). Many physical properties, such as R\'enyi entropies and partial-transpose moments related to mixed-state entanglement, can be represented as MCOs~\cite{elben2020mixed, neven_symmetry-resolved_2021,Yu_2021_Optimal,votto_2022_r2}. MCOs also yield bounds on the quantum Fisher information~\cite{Cerezo_2021_subqfi,Yu_2021_experimental,rath2021quantum} and other entanglement detection quantities~\cite{Liu_2022_detecting,rath2023barrier,rico2023entanglement} and naturally appear in the context of error mitigation~\cite{cotler2019quantum,seif2023shadow}.

A central question for the classical shadow technique, and RMs in general, concerns minimizing the number of measurements required to maintain statistical errors at a certain level. While numerous works have addressed statistical error reduction in classical shadows for single-copy observables~\cite{hadfield2020locally,hadfield2021adaptive,huang2021estimation,yen2022deterministic,vankirk2022hardware}, optimized methods for reducing statistical errors are especially vital for MCOs, where the required number of measurements typically scales exponentially with (sub-)system size~\cite{elben2022therandomized}.
In this work, we propose a framework for enhancing estimations, i.e., reducing statistical errors, for general MCOs by incorporating approximate knowledge of the quantum state of interest.
This is relevant for estimating linear ($n=1$) and non-linear ($n>1$) observables with reduced statistical errors.

Our approach is based on the technique of \emph{common random numbers}~\cite{glasserman1992some}. 
Suppose we aim to estimate the expectation value $\expp[X]$ of a random variable $X$. If we estimate $\expp[X]$ by averaging over multiple samples $X_i$, the statistical error is quantified by the variance $\Var[X]$. Now, assume we have access to a random variable $Y$, strongly correlated with $X$ \footnote{Specifically, we require that $\text{Cov}(X,Y)=\expp[XY]-\expp[X]\expp[Y]>\Var[Y]/2$, so that $\Var[X-Y]<\Var[X]$.} whose average value $\expp[Y]$ is known. We can estimate $\expp[X]$ with reduced variance $\Var[X-Y]< \Var[X]$ by averaging the random variable \mbox{$X-Y+\expp[Y]$} over \emph{commonly} sampled variables ${X_i,Y_i}$.

In this work, we employ the idea of common random numbers to introduce \emph{common randomized measurements} (CRM). Our starting point are (standard) RMs that have been experimentally performed on a quantum state $\rho$ \cite{elben2022therandomized}. To enhance the estimation of (multi-copy) observables, we utilize (approximate) knowledge of the experimental state $\rho$, provided in the form of a classically representable approximation $\sigma$,  during the classical post-processing stage. Here, $\sigma$ can  be derived from approximate theoretical modeling of the experiment or from data obtained in companion experiments. CRMs are realized by simulating classically  RMs on $\sigma$ using the same random unitaries as applied in the experiment. If $\rho$ and $\sigma$ are sufficiently close, the results of experimentally realized (on $\rho$) and simulated (on $\sigma$) RMs will be strongly correlated. Then, we can construct powerful CRM estimators for MCOs with reduced statistical error compared to the `standard' classical shadow approach. To demonstrate this, we present analytical variance bounds based on combining results on MCO~\cite{rath2021quantum,rath2023barrier} and multi-shot~\cite{seif2023shadow,zhou2022performance,helsen2022thrifty} shadow estimations, as well as two numerical examples.

{\it Randomized measurements \& classical shadows---}
Classical shadows~\cite{huang2020predicting} are classical snapshots of a quantum state that can be constructed efficiently from the experimental data acquired through RMs~\cite{elben2022therandomized}.
For concreteness, we consider here quantum systems consisting of $N$ qubits and described by a density matrix $\rho$.
RMs are generated by applying a random unitary $U$ on $\rho$, sampled from a suitable ensemble (specified below).
After applying the unitary $U$, a projective measurement on the rotated state $U\rho U^\dag$ is performed in the computational basis \mbox{${\ket{\mathbf{s}}} = {\ket{s_1, \dots , s_N}}$} with $s_i \in {0,1}$.
We assume that a total of $N_U N_M$ such RMs are performed, with $N_U$ denoting the number of sampled random unitaries $U^{(r)}$ and $N_M$ representing the number of projective measurements per random unitary.
The measurement data thus consists of $N_UN_M$ bitstrings, which we label as \mbox{$\mathbf{s}^{(r,b)}=(s_1^{(r,b)},\ldots,s_N^{(r,b)})$}, for $r=1,\dots,N_U$, and \mbox{$b=1,\dots,N_M$}.

From the measured data, one can construct $N_U$ `standard' classical shadows
\begin{equation}
    \hat{\rho}^{(r)}
    = \sum_{\s}
    \widehat{P}_\rho(\s|U^{(r)})
    \mathcal{M}^{-1}
\left(
{U^{(r)}}^\dag \ket{\s}
\bra{\s} U^{(r)}
\right),
\end{equation}
with $r=1,\dots,N_U$, and $\widehat{P}_\rho(\s|U^{(r)})=\sum_{b}\delta_{\s,\s^{(r,b)}}/N_M$
denoting the \emph{experimentally estimated outcome probabilities} of computational basis measurements performed on $U^{(r)} \rho {U^{(r)}}^\dag$. The inverse shadow channel $\mathcal{M}^{-1}$   is constructed such that, given the distribution of the random unitaries $U$, $\hat{\rho}^{(r)}$ is an unbiased estimator of $\rho$, i.e., \mbox{$ \expp[ \hat{\rho}^{(r)} ] =  \expp_U \expp_{QM}[ \hat{\rho}^{(r)} ] = \rho$}~\cite{huang2020predicting}.
Here, $ \expp_U$ denotes the average over the random unitary ensembles and $\expp_{QM}$ the quantum mechanical expectation value (for a given $U$).
While our construction of CRM shadows applies to any type of RM settings, we will consider for concreteness in the following examples Pauli measurements using random unitaries $U = \bigotimes_{i=1}^N U_i$ where each $U_i$ is uniformly
sampled in $\left\{\mathbb{1}_2,\frac{1}{\sqrt{2}}\left(\begin{smallmatrix}
    1 & 1 \\
    1 & -1
\end{smallmatrix}\right),
\frac{1}{\sqrt{2}}
\left(\begin{smallmatrix}
    1 & -i \\
    1 & +i
\end{smallmatrix}\right) \right\}$, so that  \mbox{$U_i^\dagger Z U_i=Z,X,Y$}, respectively (with $Z,X,Y$ being the Pauli matrices). 
The corresponding inverse shadow channel is such that $\mathcal{M}^{-1}(\bigotimes_i O_i)=\bigotimes_i(3O_i-\tr(O_i)\mathbb{1}_2)$~\cite{huang2020predicting}.

{\it Common randomized measurements---}
The central idea of this work is to construct  classical shadows which incorporate (approximate)  knowledge of the state $\rho$ in the form of some classically representable approximation $\sigma$. We assume that $\sigma$ is hermitian but not necessarily positive semi-definite or trace one and call it a pseudo-state for this reason. We propose building CRM shadows as
\begin{equation}
    \hat{\rho}_\sigma^{(r)}
    =
    \hat{\rho}^{(r)}
    -\sigma^{(r)}
    +\sigma ,
    \label{eq:shiftedshadow}
\end{equation}
where the term $\sigma^{(r)}$ is constructed from $\sigma$ as 
 \begin{equation}
    \sigma^{(r)}
    =
    \sum_{\s}
     {P}_\sigma(\s|U^{(r)})
    \mathcal{M}^{-1}
\left(
{U^{(r)}}^\dag \ket{\s}
\bra{\s} U^{(r)}
\right), \label{eq:def_sigma_r}
 \end{equation}
 with
 ${P}_\sigma(\s|U^{(r)})=\braket{\s|U^{(r)} \sigma {U^{(r)}}^\dag|\s}$ being the \emph{exact theoretical outcome probabilities} of (fictious) computational basis measurements on the pseudo-state $U^{(r)} \sigma {U^{(r)}}^\dag$---i.e., after $\sigma$ is rotated by the same unitary $U^{(r)}$ that has been applied in the experiment.
 Utilizing  the definition of the inverse shadow channel \cite{huang2020predicting}, we find 
\mbox{$ \expp[ {\sigma}^{(r)} ] =\expp_U [ {\sigma}^{(r)} ] = \sigma$}. Thus, $\hat{\rho}_\sigma^{(r)}$ is an unbiased estimator of $\rho$, as $
    \expp[ \hat{\rho}_\sigma^{(r)} ]
=\rho -\sigma + \sigma  = \rho$, irrespective of the choice of $\sigma$.
Crucially, the data acquisition is independent of $\sigma$, which enters only during post-processing. In particular, an optimal $\sigma$ can be chosen \emph{after} the experiment, for instance, if a new or more accurate theoretical modeling of the experiment becomes available.

The power of CRM shadows can be intuitively understood
in the limit of large numbers of measurements $N_M\gg 1$.
Then, if  $\rho \approx \sigma$, $\hat{\rho}^{(r)}\approx \rho^{(r)}$ and $ \sigma^{(r)}$ are  strongly positively correlated since they share a common source of randomness (the matrix elements of the random unitary $U^{(r)}$). Consequently, the variances of the matrix elements of $\hat{\rho}^{(r)}-\sigma^{(r)}$ are smaller than those of $\hat{\rho}^{(r)}$.
Below, we turn this intuition into rigorous performance guarantees.

We note that constructing $\sigma^{(r)}$ incurs overhead in terms of post-processing compared to standard shadow estimations.  However, as we will show below, this step can be efficiently executed (both in terms of time and memory) using suitable  representations, such as tensor networks~\cite{cirac2021matrix}. Moreover, we remark that instead of utilizing a theoretical state $\sigma$, one can build $\sigma$ from classical shadows obtained from a companion experiment that produces a state $\sigma$ close to $\rho$. This is particularly important in scenarios where a large set of RMs on a state $\sigma$ has already been acquired in such a companion experiment. This idea is presented in the supplemental material (SM, App.~D)~\cite{SM}, where we present expressions of CRM shadows that are built from the data associated with both $\rho$ and $\sigma$ and that allow for unbiased MCO estimations for $\rho$.

{\it Estimation of Pauli observables---}
We first consider estimators \mbox{$\hat{O}=\frac{1}{N_U}\sum_{r=1}^{N_U}\tr\left(O\hat \rho^{(r)}_\sigma\right)$} of expectation values $\tr(O\rho)$ of (single-copy) Pauli observables $O=\bigotimes_{i=1}^N O_i$ where each $O_i \in\{\mathbb{1}_2, X,Y,Z\}$ is a Pauli matrix. As shown in the SM~\cite{SM}, App.~B, we find for the variance of $\hat{O}$,
\begin{equation}
    \Var[\hat O]  \leq
    \frac{3^{N_A}}{N_U}
    \left(
\tr[O(\rho-\sigma)]^2
+\frac{1}{N_M}
    \right),
\end{equation}
where $N_A$ denotes the size of the support of $O$ (i.e.\ of the set $A$ of qubits $i$ where $O_i\neq \mathbb{1}_2$). 
With standard shadows, the same expression applies after replacing $\sigma$ by $0$, and our bound is consistent with Theorem~2 in Ref.~\cite{zhou2022performance}. This result demonstrates the power of CRMs:  statistical errors in estimations with classical shadows originate both from the finite number of measurement settings $N_U$ and from the finite number of experimental runs per setting $N_M$.  With CRMs, we can significantly decrease the former such that, for any value of $N_M$, the variance given by CRM shadows is smaller than the one of standard shadows if $|\tr[O(\rho-\sigma)]|\le |\tr(O\rho)|$. 
The fact that CRM shadows are useful to reduce the variance associated with finite $N_U$ is highly relevant in experiments with significant calibration times like trapped ions~\cite{brydges2019probing} or superconducting qubits~\cite{satzinger2021realizing}. Here, the number of settings $N_U$ is limited, while the value of $N_M$ can typically be taken to be large $N_M\gg1$.

{\it Estimation of MCOs with CRM shadows---} 
 Expectation values $\tr(O\rho^{\otimes n})$ of  $n$-copy observables $O$ can be estimated with (CRM) shadows employing U-statistics \cite{huang2020predicting,elben2020mixed}. Here, we use the method of `batch  shadows'~\cite{rath2023barrier} which reduces the data processing time: For an integer $m\ge n$, $m$ batch shadows $\hat{\rho}_\sigma^{[t]}$, $t=1,\dots,m$, are formed by averaging $m$ distinct groups of $N_U/m$ shadows $\hat{\rho}_\sigma^{(r)}$ (c.f., SM \cite{SM}, App.~A). We then define an estimator $ \hat O $ of  $\tr(O\rho^{\otimes n})$ as
 \begin{align}
\hat O =\frac{(m-n)!}{m!}\sum_{t_1\neq\dots \neq t_{n}}\mathrm{Tr}\left[O \left( \hat \rho_\sigma^{[t_1]}\otimes \dots\otimes \hat\rho_\sigma^{[t_n]}\right)\right].
\label{eq:hatO}
\end{align}
Since the batch shadows $\hat{\rho}_\sigma^{[t_i]}$ are statistically independent, and $\expp [\hat{\rho}_\sigma^{[t_i]}] = \rho$, then $\expp[\hat O]=\tr(O\rho^{\otimes n})$. As derived in SM~\cite{SM},  App.~B, the variance of  $\hat O$ is bounded by 
\begin{equation}
    \Var[\hat O]
    \le
    \frac{n^2 ||O_A^{(1)}||_2^2}{N_U}
    \left(
    3^{N_A}
||\rho_A-\sigma_A||_2^2
+
    \frac{2^{N_A}}{N_M}
        \right)
           +
           \mathcal{O}\!
    \left(
    \frac{1}{N_U^2}
    \right),
    \label{eq:boundMT}
\end{equation}
where  $||\cdot||_2 = \sqrt{\tr[(\cdot)^2]}$ is the Hilbert-Schmidt norm and the support $A=\textrm{supp}(O)$ of $O$ denotes a subset of $N_A$ qubits on which the MCO $O$ acts non-trivially in at least one of the copies.
Also, $\rho_A = \tr_{\bar A}(\rho)$ [$\sigma_A = \tr_{\bar A}(\sigma)$], where $\bar A$ is the complementary subset to $A$, are reduced density matrices and $O_A^{(1)}$ is an operator that acts on ($n$ copies of) $A$ while depending on $O$ and in general on $\rho$.
This  represents a key result of our work: Provided that $||\rho_A-\sigma_A||_2^2\ll ||\rho_A||_2^2$ and $N_M\gg (2/3)^{N_A} ||\rho_A||_2^{-2}$, the required number of unitaries $N_U$ is significantly reduced compared to standard shadows  [Eq.~\eqref{eq:boundMT} with \mbox{$\sigma\to 0$}]. 
Finally, we note that Eq.~\eqref{eq:boundMT}  is independent of $m$ and hence also applies  to the case of the `original' multi-copy estimators  \cite{huang2020predicting,elben2020mixed}, obtained with $m=N_U$ \cite{SM}, App.~A.

{\it Example 1: Polynomial approximations of the von Neumann entropy---}
As a first example, we consider the estimation of polynomial approximations of the von Neumann (vN) entropy \mbox{$S=-\tr(\rho_A \log \rho_A) $} of a subsystem $A$ of $N_A$ qubits, using trace moments \mbox{$p_n = \tr[\rho_A^n]$}. 
The vN entropy is an entanglement measure \cite{horodecki2009quantum} and can be used to distinguish quantum phases and transitions~\cite{eisert2010area}.
To obtain a polynomial approximation of $S$, we rewrite \mbox{$S = -\sum_\lambda \lambda \log \lambda $} expressed by the eigenvalues $\lambda$ of $\rho_A$ and perform a least-square function approximation of  $f(x)=-x\log(x)$  on in the interval $x \in (0,1)$ using polynomials of the type \mbox{$f_{n_{\max}}(x)=\sum_{n=1}^{n_{\max}} a_n \,x^n$.}
For \mbox{$n_\mathrm{max}=3$}, we obtain for instance, \mbox{$f_3(x)=137x/60-4x^2+7 x^3/4$}. Once we have obtained $f_{n_{\max}}(x)$, we build 
\begin{eqnarray}
   S_{n_{\max}}&=&\tr\left[f_{n_{\max}}(\rho_A)\right]
    =\sum_{n=1}^{n_{\max}} a_n\,p_n.
    \label{eq:Sn}
\end{eqnarray}
In the SM \cite{SM}, App.~E, we present the analytical expressions of $f_{n_{\max}}$ that show the convergence of least-square errors as $n_{\max}$ is increased and present an upper bound for the error $|S_{n_{\max}}-S|$.
 We note that, for the quantum states considered below as an illustration, our fitting procedure provides more accurate approximations  $S_{n_{\max}}$   compared to other polynomial interpolations of the same order~\cite{kontopoulou2018randomized}.

To estimate $S_{n_{\max}}$, we  rewrite each $p_n$  as an expectation value of a $n$-copy observable~\cite{ekert2002direct}, namely
\mbox{$p_n =\tr(\tau^{(n)}_A \rho_A^{\otimes n})$}, with the $n$-copy circular permutation  operator
acting as \mbox{$\tau^{(n)}_A\ket{\s^{(1)}_A}\cdots\ket{\s^{(n)}_A}=\ket{\s^{(n)}_A}\ket{\s^{(1)}_A}\cdots\ket{\s^{(n-1)}_A}$}, and use the batch shadow estimator [Eq.~\eqref{eq:hatO}] with $m=n_{\max}$ batches.
As shown in the SM \cite{SM}, App.~C, the variance bound Eq.~\eqref{eq:boundMT} for estimating $p_n$ evaluates to $O_A^{(1)}=\rho_A^{n-1}$.

\begin{figure}
\includegraphics[width=\columnwidth]{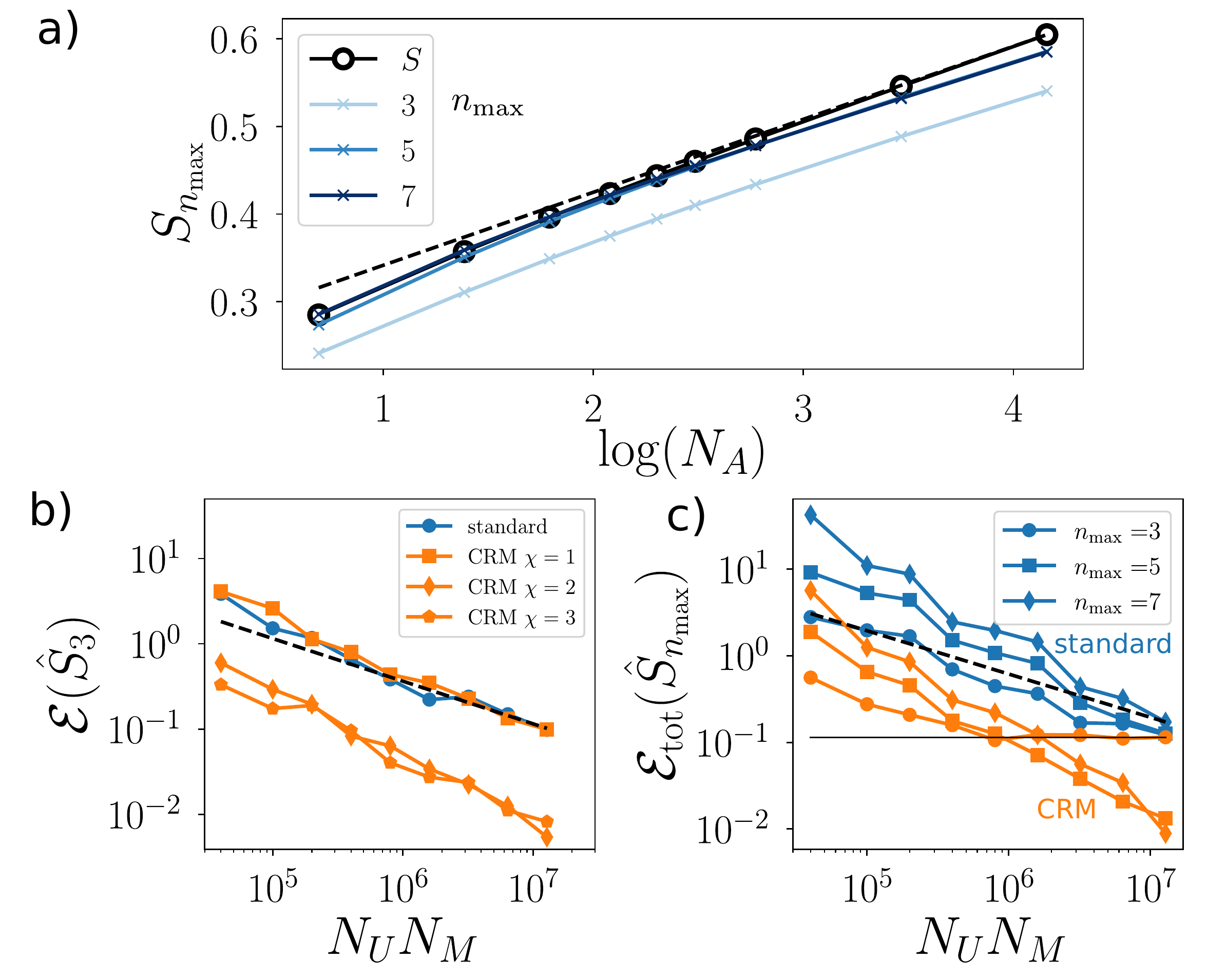}
\caption{
{\it Estimation of the von Neumann entropy in the critical Ising chain.}
a) Approximations $S_{n_{\max}}$ as a function of subsystem size $N_A=N/2$ for $n_\mathrm{max}=3,5,7$ compared to $S$.
The dashed line is a guide to the eye $\propto c/6\log(N_A)$ with $c=1/2$, the central charge of the Ising universality class~\cite{calabrese2004entanglement}.
b) Statistical relative error $\mathcal{E}( \hat{S}_3)=\expp[|\hat{S}_{3}-S_{3}|]/S_{3}$ of estimations of $S_3$, using the standard batch shadow estimation ~\cite{huang2020predicting,rath2023barrier} (blue), and the CRM method (orange) with $\sigma$ obtained from MPS approximations $\ket{\psi_\chi}$ of bond dimensions $\chi=1,2,3$.  
c) Total relative error $\mathcal{E}_{\text{tot}}( \hat{S}_{n_{\max}}) = \expp[|\hat{S}_{n_\mathrm{max}}-S|]/S$ for \mbox{$n_\mathrm{max}=3,5,7$}, and $\chi=2$. The horizontal line denotes $|S_3- S|/ S$.  In panels b) and c), we  use $N_A=N/2=8$, $N_M=1000$ is fixed, and we vary $N_U$. The black dashed lines are guides to the eye $\propto 1/\sqrt{N_UN_M}$. The average errors are obtained by averaging over $20$ and $50$ simulations for panels b) and  c), respectively. }
\label{fig:vn}
\end{figure}
As an illustration, we consider the ground state $\ket{G}$  of the critical Ising chain $
H=-\sum_{i=1}^N Z_iZ_{i+1}+X_i$ of length $N$ ($Z_i$, $X_i$ are Pauli matrices at sites $i=1,\dots,N$, and $Z_{N+1}=0$).
Since we consider the model at a critical point, the entanglement entropy $S$ of the reduced density matrix  \mbox{$\rho_A=\tr_{N/2+1,\dots,N}(\ket{G}\!\bra{G})$} of the half partition (with $N_A=N/2$ qubits) grows as  \mbox{$S=c/6\log(N_A)+\text{const}$}, where the central charge \mbox{$c=1/2$} characterizes the transition's universality class~\cite{calabrese2004entanglement}. 
In Fig.~\ref{fig:vn}a), we represent  $ S_{n_{\max}}$  as a function of $N_A=N/2$ for  different values of $n_{\max}$. Here, $\ket{G}$ is calculated from the density matrix renormalization group algorithm~\cite{schollwock2011thedensity}.
Already for $n_\mathrm{max}=3$, we observe the characteristic logarithmic scaling with $N_A$~\cite{calabrese2004entanglement}, while $n_\mathrm{max}=5,7$ provide more quantitative agreements with $S$.

We now numerically simulate a measurement of $ S_{n_{\max}}$ with `standard' classical and CRM shadows. 
In our simulations, the $N$-qubit ground state $\ket{G}$ is expressed with a Matrix-Product-State (MPS)~\cite{schollwock2011thedensity} of large bond dimension $\chi_{G}\sim 40$. We then obtain MPS approximations $\ket{\psi_\chi}$ by truncating $\ket{G}$ to much smaller bond dimensions $\chi=1,2,3$. The corresponding reduced state $\sigma$ of the first $N$ qubits is a Matrix-Product-Operator (MPO) of bond dimension $\chi^2$~\cite{schollwock2011thedensity}.
As $\chi$ increases, $\sigma$ converges to $\rho$, where we expect the optimal performances for CRM shadows.

In Fig.~\ref{fig:vn}b)-c) we show the relative statistical error $\mathcal{E}( \hat{S}_{n_{\max}})=\expp[|\hat{S}_{n_{\max}}-S_{n_{\max}}|]/S_{n_{\max}}$ as a function of  $N_UN_M$ for various $n_{\max}=3,5,7$. We  chose \mbox{$N_A=N/2=8$}, use  $N_M=1000$, and vary $N_U$. 
In panel b), we first study the behavior of  $\mathcal{E}( \hat{S}_{3})$ for $\chi=1,2,3$. For $\chi=1$, the approximation $\sigma$ corresponds to a product state, which is too inaccurate to obtain any improvement with CRM shadows over standard shadows. For $\chi=2,3$ instead, the approximation $\sigma_\chi$ is sufficiently accurate to significantly decrease the statistical errors. 
In panel c), we study the \emph{total} relative error
$\mathcal{E}_{\text{tot}}( \hat{S}_{n_{\max}})=\expp[|\hat{S}_{n_{\max}}-S|]/S$. This error includes statistical errors in estimating ${S}_{n_\mathrm{max}}$, but also the systematic error $|S_{n_\mathrm{max}}-S|$.
For small values of $N_UN_M$, where statistical errors dominate, the error increases with increasing $n_{\mathrm{max}}$ [which we  attribute to the prefactor $n^2$ in the variance bound Eq.~\eqref{eq:boundMT}].
At large numbers of measurements, the error saturates to the systematic error $|S_{n_\mathrm{max}}-S|/S$ (only visualized here for $n_{\max} = 3$, black line), and it becomes more advantageous to use larger values of $n_{\max}$.

{\it Example 2: Fidelity estimation---}
Our second example focuses on single-copy observables ($n=1$); specifically, we consider direct fidelity estimation~\cite{flammia2011direct,dasilva2011practical,cerezo2020variationalquantum}. Here the goal is to estimate the fidelity $\mathcal{F}_\psi =\braket{\psi|\rho |\psi} $ between the prepared quantum state  $\rho$ and a pure theoretical state $\ket{\psi}$, i.e., we take $O$ to be the projector $O=\ket{\psi}\!\bra{\psi}$.

Our motivation for enhanced CRM fidelity estimates is two-fold: Firstly, fidelity estimation allows us to certify the preparation of a quantum state within a quantum device. However, while $\mathcal{F}_{\psi}$ can be efficiently estimated with (standard) classical shadows constructed from \emph{global} Clifford measurements \cite{huang2020predicting}, fidelity estimation can be challenging with (standard) \emph{local} RMs due to a potential exponential scaling of the required number of measurements \cite{huang2020predicting}.
Secondly, fidelity estimation can also be used to identify suitable CRM priors $\sigma$ for estimating other MCOs:  Direct inspection of Eq.~\eqref{eq:boundMT} indeed reveals that CRM shadows provide lower variance compared to standard shadows when $\mathcal{F}_{\phi}\ge 1/2$ (considering for simplicity an MCO $O$ with full support ($N_A=N$) and a pure state prior $\sigma=\ket{\phi}\!\bra{\phi}$).

We  propose an iterative procedure to find useful priors for CRM shadows as follows:
(i) Starting with a prior $\sigma=\ket{\phi}\!\bra{\phi}$, we estimate $\mathcal{F}_{\phi}$ using either CRM shadows $\hat{\rho}^{(r)}_\sigma$ or standard shadows $\hat{\rho}^{(r)}$. The choice can be made during post-processing by comparing empirical variances, as illustrated in the numerical example below.
(ii) If $\mathcal{F}_{\phi}\leq \mathsf{F}$ falls below a specific threshold $\mathsf{F}\geq 1/2$, we define a new prior, which may involve more classical computation. We then repeat step (i). Once we have found a prior $\sigma = \ket{\phi}\!\bra{\phi}$ characterized by a sufficiently high fidelity $\mathcal{F}_{\phi}$,  we can  perform enhanced estimations on arbitrary MCOs $O$. This includes fidelities $\mathcal{F}_\psi$ to any other quantum state $\ket{\psi}$. Performance guarantees are provided by Eq.~\eqref{eq:boundMT} with the measured value of  $\mathcal{F}_{\phi}$. Importantly, the entire iterative procedure can be conducted on a single RM dataset, as the choice of the prior $\sigma$ is only incorporated during the post-processing stage.  This is in contrast with importance sampling methods~\cite{flammia2011direct,dasilva2011practical}, where the choice of measurement settings for data acquisition depends on the prior.

\begin{figure}
    \centering
    \includegraphics[width=\columnwidth]{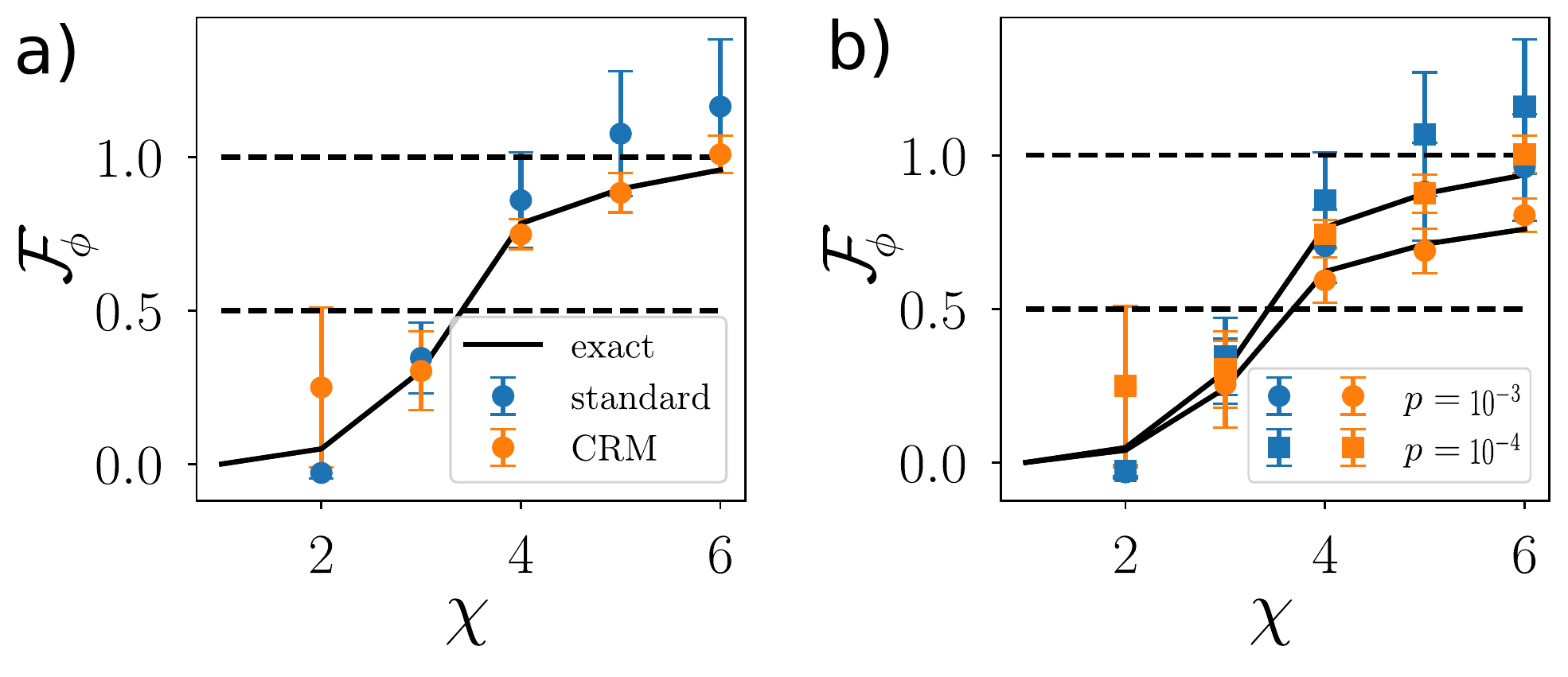}
    \caption{{\it Fidelity estimation  in (noisy) random quantum circuits ---} 
    Panel a) shows the estimated fidelities $\widehat{\mathcal{F}}_\phi$ of the prepared state $\rho$ and the theoretical prior states $\sigma = \ket{\phi}\!\bra{\phi}$ as a function of their bond dimension $\chi$.  Here, $\rho$ is a $N= 30$-qubit pure state generated from an ideal noiseless ($p=0$) random quantum circuit of depth $d =6$ and $\ket{\phi}$ are obtained by truncating $\rho$ to bond dimension $\chi$.
    In panel b), each gate in the circuit is perturbed by local depolarization noise with strength $p$ resulting in a mixed state $\rho$. The prior state $\sigma$ is the same as in a). 
     For both panels, we compare CRM estimation (orange dots) with standard shadow estimation (blue dots). We fix $N_U = 15$ and $N_M =10^5$. The error-bars are evaluated as standard errors of the mean over random unitaries.
     The black solid lines denote the exact fidelity
     $\mathcal{F}_{\phi}$. 
     The black dashed lines are guides to the eye for $0.5$ and $1$ respectively.}
    \label{fig:fidelity}
\end{figure}

As a numerical example, let us consider a state $\rho$, which is prepared  with a uni-dimensional random circuit composed of $d$ alternating layers of single and neighboring two-qubit Haar-random gates. Each gate is subject to local depolarization noise with probability $p$~\cite{pastaq}.
We then numerically simulate the RMs occurring in the experiment. 
In our numerical experiment, CRM priors $\sigma=\ket{\phi}\!\bra{\phi}$ correspond to  MPSs $\ket{\phi}$ with bond dimensions $\chi$ obtained from truncating  the exact output state of the noiseless quantum circuit.
Note that with these priors, CRM estimations can be computed in $\mathrm{poly}(\chi)$ time in the MPS formalism~\cite{schollwock2011thedensity}. 
As $\chi$ increases, the fidelity ${\mathcal{F}}_\phi$ increases, but the computational cost in estimating $\widehat{\mathcal{F}}_\phi$ with CRM shadows also grows. 
Fig.~\ref{fig:fidelity} shows the  estimations $\widehat{\mathcal{F}}_\phi$ for a $N=30$-qubit noiseless [$p=0$, panel a)] and noisy state [$p=10^{-3},10^{-4}$, panel b)], with error bars calculated as the standard error of the mean over random unitaries. 
When $\chi$ increases, the estimated fidelity $\mathcal{F}_\phi$ increases, and the error bars of the CRM estimations decrease as the CRM shadows become more accurate.
At small $\chi$ instead, the CRM shadows fail to provide improved estimations and have larger error bars compared to (standard) classical shadows as seen in Fig.~\ref{fig:fidelity}a).
These features are similarly observed in the case of the noisy experimental state in Fig.~\ref{fig:fidelity}b), where the pure state $\ket{\phi}$ remains always different from the mixed state $\rho$.

{\it Conclusion and outlook---}
CRM shadows provide a readily applicable tool to significantly  enhance  the estimation of linear and multi-copy observables by incorporating approximate knowledge of the quantum state of interest in the post-processing of RM experiments. 
Besides the presented examples, we envision a wide range of applications, from gradient estimation in variational quantum algorithms \cite{sack2022avoiding} to the probing of quantum phases of matter \cite{huang2022provable,lewis2023improved,onorati2023efficient}.
For future work, it would be interesting to study the potential benefit of using, in addition to our method, importance sampling or adaptive techniques such as the one developed to access the purity $p_2$ with RM~\cite{rath2021importance}, or improved post-processing methods from measurements obtained using auxiliary systems~\cite{tran2022measuring}.

\begin{acknowledgments}
We thank Daniel K.~Mark and Steve Flammia for helpful discussions and Richard Kueng and Hsin-Yuan (Robert) Huang for their valuable comments on our manuscript.
Work in Grenoble is funded by the French National Research Agency via the JCJC project QRand (ANR-20-CE47-0005), and via the research programs EPIQ (ANR-22-PETQ-0007, Plan France 2030), and QUBITAF (ANR-22-PETQ-0004, Plan France 2030).
B.V.\ acknowledges funding from the Austrian Science Foundation (FWF, P 32597 N).
A.R.\ acknowledges support by Laboratoire d'excellence LANEF in Grenoble (ANR-10-LABX-51-01) and from the Grenoble Nanoscience Foundation. B.S.\ is supported by Caltech Summer Undergraduate Research Fellowship (SURF).
J.P.\ acknowledges funding from  the U.S.\ Department of Energy Office of Science, Office of Advanced Scientific Computing Research, (DE-NA0003525, DE-SC0020290), the U.S. Department of Energy Quantum Systems Accelerator, and the National Science Foundation (PHY-1733907). The Institute for Quantum Information and Matter is an NSF Physics Frontiers Center. 
A.E.\ acknowledges funding by the German National Academy of Sciences Leopoldina under the grant number LPDS 2021-02 and by the Walter Burke Institute for Theoretical Physics at Caltech.

Our Julia code uses ITensor~\cite{fishman2022theitensor} and PastaQ~\cite{pastaq}, and is available \href{https://github.com/bvermersch/RandomMeas.jl}{here}.
\end{acknowledgments}
%

\clearpage
\onecolumngrid
\appendix

\begin{center}
{\large \textbf{  Supplemental Material: Enhanced estimation of quantum properties with common randomized measurements}}
\end{center}

\section{Statistical analysis of  CRM shadows}
In this appendix, we provide a general variance bound for  estimating (multi-copy) observables with CRM shadows. 

\subsection{General variance formula}
In this subsection, we recapitulate the variance of the batch shadow estimator $\hat{O}$, defined in Eq.~\eqref{eq:hatO} of the main text (MT), as derived in Ref.~\cite{rath2023barrier} and adapt it to the case of CRM shadows.  
The batch shadows that appear in Eq.~\eqref{eq:hatO} are defined as~\cite{rath2023barrier}
 \begin{equation}
    \hat{\rho}_\sigma^{[t]}
    =
   \frac{N_U}{m}
   \sum_{1+(t-1)(N_U/m)}
   ^{t(N_U/m)}\hat{\rho}_\sigma^{(r)}
   \label{eq:batch}
\end{equation}
where the CRM shadows $\hat{\rho}_\sigma^{(r)}$ are defined in Eq.~\eqref{eq:shiftedshadow} of the MT. Here,  $t=1,\dots,m$ with $m\ge n$ and we assume for simplicity that $m$ divides $N_U$.
An unbiased estimator of $\tr[O \rho^{\otimes n}]$ is then given by the U-statistic of batch shadows~\cite{Hoeffding1992} (see also Refs.~\cite{huang2020predicting,elben2020mixed})
\begin{align}
\hat O =\frac{(m-n)!}{m!} \sum_{\substack{t_1\neq t_2\neq\dots \neq t_{n} \\ t_i \in \{1, \dots, m\}}}\mathrm{Tr}\left[O \left( \hat \rho_\sigma^{[t_1]}\otimes\hat \rho_\sigma^{[t_2]}\otimes\dots\otimes \hat\rho_\sigma^{[t_n]}\right)\right]
\label{eq:hatO_SM}.
\end{align}
As stated in the main text, since the batch shadows $\hat{\rho}_\sigma^{[t_i]}$ are statistically independent, and $\expp [\hat{\rho}_\sigma^{[t_i]}] = \rho$, it follows that $\expp[\hat O]=\tr(O\rho^{\otimes n})$ . Let us mention three relevant limiting cases for our analysis: For $\sigma=0$, we recover estimations with standard (batch) shadows~\cite{rath2023barrier}. 
For $\sigma=0$, and $m=N_U$, $\hat O$ coincides with the standard shadow estimator for MCO presented in Refs.~\cite{huang2020predicting,elben2020mixed}. 
Finally, for linear observables $n=1$, the notion of batch shadows becomes meaningless as the two averages in Eq.~\eqref{eq:batch} and Eq.~\eqref{eq:hatO_SM} can be combined, i.e., the estimation does not depend on $m$ anymore. 

The variance of the batch shadow estimator $\hat O $ has been calculated in Ref.~\cite{rath2023barrier} for the case of `standard classical shadows'. As this derivation only relies on the condition $\expp[\hat \rho^{(r)}]=\rho$, the same result applies for CRM shadows \cite[Eq.~(C27)]{rath2023barrier}:
\begin{align}
        \Var[\hat{O}]
    &=  \sum_{\ell=1}^n \frac{\binom{n}{\ell}^2}{\binom{m}{\ell}} \left(\frac{m}{N_U}\right)^\ell \Big[\sum_{k=1}^\ell \binom{\ell}{k} (-1)^{\ell-k} \Var_k \Big] = \frac{n^2}{N_U} \Var_1 + \frac{n^2(n-1)^2\frac{m}{m-1}}{2N_U^2} (\Var_2 - 2\Var_1) + \mathcal{O}\Big(\frac{1}{N_U^3}\Big),
     \label{eq:def_V}
\end{align}
with the terms 
\begin{align}
    \Var_k = \Var
    \Bigg[
    \mathrm{Tr}
    \left[ O_\mathrm{sym}
    \left(\bigotimes_{r = 1}^k \hat{\rho}_\sigma^{(r)} \otimes\rho^{\otimes(n-k)} \right)
    \right] \Bigg]
    =\Var
        \Bigg[
    \mathrm{Tr}
    \left( O^{(k)}
    \bigotimes_{r = 1}^k \hat{\rho}_\sigma^{(r)} 
    \right) \Bigg] \quad \text{for} \quad k=1,\dots, n
    , \label{eq:def_Vk}
\end{align}
depending on the CRM shadows $ \hat{\rho}_\sigma^{(r)}$. Here, we defined a symmetrized $n$-copy operator 
$O_\mathrm{sym}=\frac{1}{n!}\sum_{\pi \in \mathcal{S}_n} W_\pi^\dagger OW_\pi$ and its $\rho$-dependent partial traces $O^{(k)}=\tr_{k+1,\dots,n}\big[O_\mathrm{sym} \big(\mathbb{1}_{2^N}^{\otimes k}\otimes\rho^{\otimes (n-k)}\big)\big]$.
The operators $W_\pi$ are $n$-copy permutation operators, with $\pi=(\pi(1),\dots,\pi(n))$, acting as $W_\pi(\otimes_i \ket{\s_i})=\otimes_i \ket{\s_{\pi(i)}}$, and $\mathcal{S}_n$ denotes the symmetric group.

For later use, we define the support $\textrm{supp}(O)$ of a multi-copy operator $O$ as the subpartition of the quantum system $\mathcal{S}$ on which $O$ acts non-trivially in at least one of their copies. Then, by definition of $O^{(k)}$, $\textrm{supp}(O^{(k)}) \subseteq \textrm{supp}(O)$.
Denoting $A\equiv \textrm{supp}(O)$, we can factorize $O^{(k)}= O^{(k)}_{ A}\otimes \mathbb{1}^{\otimes k}_{\bar A}$ with $O^{(k)}_{A} = \tr_{k+1,\dots,n}\big[(O_A)_\mathrm{sym}\big(\mathbb{1}_{A}^{\otimes k}\otimes\rho_A^{\otimes (n-k)}\big)\big]$ and $(O_A)_\mathrm{sym}$ the symmetrization of the restriction $O_A$ of $O$ to $A$.

\subsection{Leading order term $(n^2/N_U)\Var_1$}
    We evaluate now the leading order term $\frac{n^2}{N_U} \Var_1$ in Eq.~\eqref{eq:def_V}.
    It is dominant in the limit  $N_U\to \infty$, and solely determines the variance of the estimation of standard single-copy observables [$n=1$ in Eq.~\eqref{eq:def_V}] as higher order terms in Eq.~\eqref{eq:def_V}  vanish. Interestingly, this  term does not depend on the number of batches $m$, so we typically choose a minimal number  $m=n$ of batch shadows to evaluate Eq.~\eqref{eq:hatO_SM}, as it leads to  a minimal postprocessing effort: $\mathcal{O}(m^n)$ terms have to be evaluated in Eq.~\eqref{eq:hatO_SM}.

    Here, and in the following,  we use the fact that the inverse shadow channel $\mathcal{M}^{-1}$ used to define $\hat{\rho}_\sigma^{(r)}$ is Hermitian-preserving and self-adjoint (c.f., Ref.~\cite{huang2020predicting} to see that this is necessarily the case for shadows built from randomized measurements). Using this, we first evaluate
\begin{align}
\tr(O^{(1)} \hat \rho_\sigma^{(r)})
&=
\sum_{\s}
\left(
\widehat{P}_\rho(\s|U)
-{P}_\sigma(\s|U)
\right)
\tr
\left[
O^{(1)}
\mathcal{M}^{-1}
\left(
U^\dag \ket{\s}
\bra{\s} U
\right)
\right] + \tr(O^{(1)}\sigma)
\nonumber \\
&=
\sum_{\s}
\left(
\widehat{P}_\rho(\s|U)
-{P}_\sigma(\s|U)
\right)
[\mathcal{M}^{-1}
(O^{(1)})](U,\s)+ \tr(O^{(1)}\sigma),
\end{align}
where we dropped the label ${(r)}$ and introduced the short-hand notation $\tr
\left[
\mathcal{M}^{-1}
\left(O^{(1)}\right)
U^\dag \ket{\s}
\bra{\s} U
\right] = [\mathcal{M}^{-1}
(O^{(1)})](U,\s)$.  
With this, we obtain
\begin{align}
    \Var_1
    =&\Var
    [\tr(O^{(1)} \hat \rho_\sigma^{(r)})]
\nonumber \\
=&\Var
    \left[
    \sum_{\s}
    \left(
    \widehat{P}_\rho(\s|U)
    -
    {P}_\sigma(\s|U)
    \right)
[\mathcal{M}^{-1}
(O^{(1)})](U,s)
    \right]
    \nonumber \\
    =&
    \expp_U
    \left[
     \sum_{\s,\s'}
    \expp_{QM}\left[
    \left(
    \widehat{P}_\rho(\s|U)
     -
   {P}_\sigma(\s|U)
    \right)
    \left(
    \widehat{P}_\rho(\s'|U)
     -
    {P}_\sigma(\s'|U)
    \right)
    \right]
[\mathcal{M}^{-1}
(O^{(1)})](U,\s)
[\mathcal{M}^{-1}
(O^{(1)})](U,\s')
    \right]
    \nonumber \\
    &-\tr\left[O^{(1)} (\rho-\sigma)\right]^2.
    \label{eq:var_step}
\end{align}
Now we use that (see, e.g., Refs.~\cite{Vermersch_2018_PRA,elben2019statistical})
\begin{equation}
\expp_{QM}[\widehat{P}_\rho(\s|U)\widehat{P}_\rho(\s'|U)]
=P_\rho(\s|U)P_\rho(\s'|U)
+\frac{\delta_{\mathbf{s},\mathbf{s}'} P_\rho(\s|U)
-P_\rho(\s|U)P_\rho(\s'|U)}
{N_M} .
\end{equation}
Inserting this into Eq.~\eqref{eq:var_step}, we find
\begin{align}
    \Var_1
    &=
    \Var_U
    [f_{\rho,\sigma}(U)]
    +\frac{\expp_U[g_{\rho}(U)]}{N_M}
    \label{eq:var_step2}
\end{align}
with $\Var_U$ and $\expp_U$ denoting  variance and expectation with respect to only the distribution of the unitaries $U$ (as we have performed the quantum mechanical average explicitly). The functions can be written explicitly as
\begin{align}
    f_{\rho,\sigma}(U)
    &=    \sum_{\s}
        \big(
   P_\rho(\s|U)
   -P_\sigma(\s|U) \big)
[\mathcal{M}^{-1}
(O^{(1)})](U,\s)
\end{align}
and 
\begin{align}
    g_{\rho}(U)
    &=
    \sum_{\s}
           P_\rho(\s|U) \Big( [\mathcal{M}^{-1}(O^{(1)})](U,\s) \Big)^2
    -\Big(\sum_{\s}
   P_\rho(\s|U)
[\mathcal{M}^{-1}
(O^{(1)})](U,\s)\Big)^2.
\end{align}
We remark that the $N_M$-independent first term in Eq.~\eqref{eq:var_step2} depends on both $\rho$ and $\sigma$ and vanishes for $\rho=\sigma$ ($f_{\rho,\rho} = 0$ for any $\rho$). It accounts for the variance of $\hat{O}$ due to a finite number $N_U$ of random unitaries $U$ and is present even for $N_M\to \infty$. In contrast, the  $N_M$-dependent second term in Eq.~\eqref{eq:var_step2}  quantifies the average quantum shot  noise arising from a finite number $N_M$ of computational basis measurements per random unitary and depends only on the experimental state $\rho$. 

Combining these results and inserting them into Eq.~\eqref{eq:def_V}, we obtain  
\begin{align}
    \Var(\hat O)
       &=
    \frac{n^2}{N_U}
    \left(
    \Var_U
    [f_{\rho,\sigma}(U)]
    +
   \frac{
\expp_U[g_\rho(U)]
    }
    {N_M}
\right)
+\mathcal{O}\left(
\frac{1}{N_U^2}
\right)
\label{eq:firstbound}
\end{align}
Note that although it is difficult to bound the higher order term  $\mathcal{O}(1/N_U^2)$ explicitly, we have the guarantee that this term decays to zero when $N_M\to\infty$ and $\sigma \to \rho$.  In this case, the CRM shadows involved in the estimator $\hat{O}$ satisfy $\hat \rho^{(r)}\to\sigma^{(r)}$ and therefore $\hat \rho^{(r)}_\sigma\to\sigma$ becomes constant.

\section{Variance bounds for local Pauli measurements}
The analysis of the previous appendix applies to all unitary ensembles that can be used to define classical shadows (i.e., give rise to a tomographically complete set of measurements). Here, we consider Pauli measurements realized by local random unitaries of the form $U=U_1\otimes \dots \otimes U_N$, where each $U_i$ is sampled independently from the set $\mathcal{U}=\{\mathbb{1}_2,\frac{1}{\sqrt{2}}\left(\begin{smallmatrix}
    1 & 1 \\
    1 & -1
\end{smallmatrix}\right),
\frac{1}{\sqrt{2}}
\left(\begin{smallmatrix}
    1 & -i \\
    1 & +i
\end{smallmatrix}\right) \}$, so that  $U_i^\dagger Z U_i=Z,X,Y$, respectively.
In this case, the inverse shadow channel used to define $\hat{\rho}_\sigma$ [Eq.~\eqref{eq:shiftedshadow} in the MT] reads as 

\begin{align}
    \mathcal{M}^{-1}
    \Big(\bigotimes_i O_i\Big)
    &=\bigotimes_i \mathcal{M}_i^{-1}
    (O_i)
\quad \text{with} \quad
\mathcal{M}_i^{-1}(O_i)=
3O_i-\mathbb{1}_2
\tr(O_i),
\label{eq:localISC}
\end{align}
for product observables $O=\bigotimes_i O_i$ and one can extend this definition to non-product operators by linearity \cite{huang2020predicting}.

\subsection{Estimating single-copy Pauli observables}
We first consider single-copy ($n=1$) observables $O$. In this case, the operator $O^{(1)}$ defined below Eq.~\eqref{eq:def_Vk} evaluates to $O^{(1)}=O$. In addition, we specify $O$ to be a
Pauli string  of the form
$O=\gamma=\bigotimes_{i=1}^N \gamma_i$ with $\gamma_i \in\{\mathbb{1}_2, X,Y,Z\}$ and $X,Y,Z$.  We denote with $A=\textrm{supp}(O)$ the subset of $N_A$ qubits  where $O$ acts non-trivially, i.e., $\gamma_i\neq \mathbb{1}_2$ for $i\in A$ and $\gamma_i= \mathbb{1}_2$ for $i\in \bar{A}$. We define $U_A=\bigotimes_{i\in A}U_i$.

We bound $\Var(\hat O)$ starting from the expression Eq.~\eqref{eq:firstbound}. 
First, it follows directly from Eq.~\eqref{eq:localISC} that
$\mathcal{M}^{-1}(\gamma)=3^{N_A} \gamma$. Introducing the unitary $V_\gamma=\bigotimes_{i \in A} V_i$, with $V_i  \in \mathcal{U}$ such that $\gamma_i =V_i^\dag Z V_i $ for all $i\in A$,  we can rewrite
\begin{equation}
\gamma(U,\s)\equiv \tr
\left[
\gamma 
U^\dag \ket{\s}
\bra{\s} U
\right] = \delta_{U_A,V_\gamma}\braket{\s|Z_A|\s}
\end{equation}
with $Z_A=\bigotimes_{i}(Z\delta_{i\in A}+\mathbb{1}_2\delta_{i \notin A})=\sum_{\s}(-1)^{\sum_{i \in A} s_i}\ket{\s}\!\bra{\s}$. With these definitions,  we 
rewrite the first term in Eq.~\eqref{eq:firstbound} as 
\begin{align}
    f_{\rho,\sigma}(U) = 3^{N_A} \sum_{\s}  \big(P_\rho(\s|U)-P_\sigma(\s|U)\big) \gamma(U,\s)
    &=
    3^{N_A} \sum_{\s}  \big(P_\rho(\s|U)-P_\sigma(\s|U)\big) \delta_{U_A,V_\gamma}\braket{\s|Z_A|\s}
    \nonumber \\
    &=3^{N_A} \delta_{U_A,V_\gamma}
    \tr
    \left(
    (\rho-\sigma)U^\dag Z_A U
    \right)\nonumber\\&=3^{N_A}\delta_{U_A,V_\gamma}\tr((\rho-\sigma)\gamma), 
\end{align}
where we used that $Z_A$ commutes with each $\ket{\s}\!\bra{\s}$. We obtain
\begin{align}
\Var_U[f_{\rho,\sigma}(U)]
= 9^{N_A} \tr((\rho-\sigma)\gamma)^2 \, \Var_U[\delta_{U_A,V_\gamma}] = (3^{N_A}-1) \, \tr((\rho-\sigma)\gamma)^2 
\end{align}
using the fact that $\Var_U[\delta_{U_A,V_\gamma}] = \expp_U[\delta_{U_A,V_\gamma}^2] - \expp_U[\delta_{U_A,V_\gamma}]^2 = 1/3^{N_A} - 1/9^{N_A}$ irrespective of $\gamma$. Similarly, we can proceed with the second term in 
Eq.~\eqref{eq:firstbound}
\begin{align}
    g_{\rho}(U)&=
    \sum_{\s}
P_\rho(\s|U) \mathcal{M}^{-1}(\gamma)(U,\s)^2
    -\left(
     \sum_{\s}
P_\rho(\s|U) \mathcal{M}^{-1}(\gamma)(U,\s)
\right)^2
\nonumber \\
&=
9^{N_A}
    \left(
    \sum_\s P_\rho(\s|U) \gamma(U,\s)^2
    \right)
    -
    9^{N_A}\delta_{U_A,V_\gamma}\tr(\rho\gamma)^2
    =9^{N_A} \delta_{U_A,V_\gamma}(1-\tr(\rho\gamma)^2),
\end{align}
and thus 
\begin{align}
E_U[g_{\rho}(U)]
&=3^{N_A} (1-\tr(\rho\gamma)^2).
    \end{align}
    Inserting into Eq.~\eqref{eq:def_V} and recalling that higher orders terms $\mathcal{O}(1/N_U^2)$ are absent for linear observables, we find
\begin{equation}
    \Var(\hat O)
    =
    \frac{1}{N_U}
    \left(
    (3^{N_A}-1)
\tr(O(\rho-\sigma))^2
+\frac{3^{N_A} (1-\tr(\rho\gamma)^2)}{N_M}
    \right)
    \le 
        \frac{3^{N_A}}{N_U}
    \left(
\tr[O(\rho-\sigma)]^2
+\frac{1}{N_M}
    \right)
\end{equation}
as stated in the MT.

\subsection{Estimating general MCOs}

We now bound the variance Eq.~\eqref{eq:firstbound} for a general multi-copy observable $O$ with corresponding Hermitian operator $O^{(1)}$ defined below in Eq.~\eqref{eq:def_Vk}. We denote the support of $O$ with $A=\textrm{supp}(O)\supseteq \textrm{supp}(O^{(1)})$, such that, up to relabeling of the qubits, we can write, $O^{(1)}=O^{(1)}_A\otimes \mathbb{1}_{\bar A}$. We write
$O^{(1)}_A$ in the basis of  the $4^{N_A}$  Pauli strings $\gamma_A$, $O^{(1)}_A=\frac{1}{2^{N_A}}\sum_{\gamma_A} \tr(O^{(1)}_A\gamma_A)\gamma_A$ (where the Pauli strings are orthogonal: $\tr(\gamma_A \gamma_{A}')=2^{N_A}\delta_{\gamma_A,\gamma_A'}$), 
and first note that 
\begin{align}
    [\mathcal{M}^{-1}
(O^{(1)})](U,\s)
&=\tr\left[\mathcal{M}^{-1}
(O^{(1)}) U^\dag \ket{\s}
\bra{\s} U \right] \nonumber\\
&=\tr\left[\mathcal{M}_A^{-1}
(O_A^{(1)})  U_A^\dag \ket{\s_A}
\bra{\s_A} U_A \right] \nonumber \\
&=\frac{1}{2^{N_A}}\sum_{\gamma_A}3^{N_{\Gamma}}\tr(O^{(1)}_A\gamma_A)\gamma_{A}(U_A,\s_A),
\end{align}
with $\Gamma = \textrm{supp}(\gamma_A) \subseteq A$ denoting the support of $\gamma_A$ consisting of $N_{\Gamma}$ qubits. 
This implies that  $f_{\rho,\sigma}(U)$ depends only on  reduced quantites acting on $A$ only:
\begin{align}
     f_{\rho,\sigma}(U)
    &=    \sum_{\s}
        \big(
   P_\rho(\s|U)
   -P_\sigma(\s|U) \big)
[\mathcal{M}^{-1}
(O^{(1)})](U,\s)  \nonumber\\
&=    \sum_{\s_A}
        \big(
   P_{\rho_A}(\s_A|U_A)
   -P_{\sigma_A}(\s_A|U_A) \big)
\left(
\frac{1}{2^{N_A}}\sum_{\gamma_A}3^{N_{\Gamma}}\tr(O_A^{(1)}\gamma_A)\gamma_{A}(U_A,\s_A)
\right)
\end{align}
with the reduced density matrices $\rho_A=\tr_{\bar A}(\rho),\sigma_A=\tr_{\bar A}(\sigma)$. 
We now use the Cauchy-Schwartz inequality
\begin{align}
f_{\rho,\sigma}(U)^2
\le
\left(
 \sum_{\s_A}
 \left[
   P_{\rho_A}(\s_A|U_A)
   -P_{\sigma_A}(\s_A|U_A)
   \right]^2
   \right)
\left(
 \sum_{\s_A}
 \left[
\frac{1}{2^{N_A}}\sum_{\gamma_A}3^{N_{\Gamma}}\tr(O_A^{(1)}\gamma_A)\gamma_{A}(U_A,\s_A)
\right]^2
\right).
\end{align}
The first factor can be bounded as 
\begin{align}
\sum_{\s_A} \big(P_{\rho_A}(\s_A|U_A)-P_{\sigma_A}(\s_A|U_A)\big)^2
&= \sum_{\s_A} \braket{\s_A| U_A(\rho_A-\sigma_A)U_A^\dag|\s_A}^2
\le \sum_{\s_A} \braket{\s_A| [U_A(\rho_A-\sigma_A)U_A^\dag]^2|\s_A}
= ||\rho_A-\sigma_A||_2^2
\end{align}
where we have used that $\rho_A-\sigma_A$ is Hermitian.
As before, we denote by $V_{{\gamma}} = \bigotimes_{i\in \Gamma} V_i$, $V_i \in \mathcal{U}$, the unitary that maps $\gamma_i$ to $Z_i$ for all $i\in \Gamma$. Further, we define $Z_{\Gamma}=\bigotimes_{i \in A}(Z\delta_{i\in \Gamma}+\mathbb{1}_2\delta_{i \notin \Gamma})$, such that $\gamma(U_A,\s_A)=\braket{\s_A|Z_{\Gamma}|\s_A}\delta_{U_{\Gamma},V_{\gamma}}$; and analogously for $\gamma'_A$, $V_{{\gamma'}}$ and $Z_{\Gamma'}$. We then have 
\begin{align}
\sum_{\s_A}
\gamma_A(U_A,\s_A)
\gamma'_A(U_A,\s_A)
&=
\sum_{\s_A} \braket{\s_A|Z_{\Gamma}|\s_A}\braket{\s_A|Z_{\Gamma'}|\s_A}
\delta_{U_{\Gamma},V_{\gamma}}
\delta_{U_{\Gamma'},V_{\gamma'}} \nonumber \\
&=
\prod_i
\left(
1+(-1)^{\delta_{i\in {\Gamma}}+\delta_{i\in {\Gamma'}}}
\right)
\delta_{U_{\Gamma},V_{\gamma}}
\delta_{U_{\Gamma'},V_{\gamma'}}
\nonumber \\
&=\prod_i
\left(
2[\delta_{i \in {\Gamma}}\delta_{i \in {\Gamma'}}+
\delta_{i \notin {\Gamma}}\delta_{i \notin {\Gamma'}}]
\right)
\delta_{U_{\Gamma},V_{\gamma}}
\delta_{U_{\Gamma'},V_{\gamma'}}
\nonumber \\
&
=2^{N_A}\delta_{{\Gamma},{\Gamma'}}\delta_{U_{\Gamma},V_{\gamma}}\delta_{U_{\Gamma'},V_{\gamma'}}\nonumber \\
&=2^{N_A}\delta_{\gamma_A,\gamma_A'}\delta_{U_{\Gamma},V_{\gamma}}
\end{align}
where we have used in the last equality that two Pauli strings that have the same support $\Gamma=\Gamma'$ and that are mapped to $Z_{\Gamma}=Z_{\Gamma'}$ via the same transformation $U_A$ are necessarily equal.
Hence we get
\begin{align}
f_{\rho,\sigma}(U)^2
\le
\frac{1}{2^{N_A}}
 ||\rho_A-\sigma_A||_2^2
\sum_{\gamma_A}
\delta_{U_{\Gamma},V_{\gamma}}
9^{N_{\Gamma}}\tr(O_A^{(1)}\gamma_A)^2.
\end{align}

With $\expp_U[\delta_{U_{\Gamma},V_{\gamma}}]=1/3^{N_{\Gamma}}$, we obtain 
\begin{align}
\Var_U[f_{\rho,\sigma}(U)^2]\le
    \expp_U
    [f_{\rho,\sigma}(U)^2
    ]
    &\le 
    \frac{ ||\rho_A-\sigma_A||_2^2 }{2^{N_A}}
    \sum_{\gamma_A}3^{N_{\Gamma}}
    \tr(O_A^{(1)}\gamma_A)^2 \nonumber \\
&\le 
    3^{N_A} \, \frac{ ||\rho_A-\sigma_A||_2^2 }{2^{N_A}}
    \sum_{\gamma_A }   \tr(O_A^{(1)}\gamma_A)^2  \nonumber  \\
    &=  3^{N_A} \,
  ||O_A^{(1)}||_2^2\, ||\rho_A-\sigma_A||_2^2.
\end{align}
where we have used that $||O^{(1)}_A||_2^2=\tr([O_A^{(1)}]^2)=\sum_{\gamma_A}\tr(O_A^{(1)}\gamma_A)^2/2^{N_A}$ (which can be easily proven using $\tr(\gamma_A \gamma_A')=2^{N_A}\delta_{\gamma_A,\gamma_A'}$).

In order to bound the second term in Eq.~\eqref{eq:firstbound}, we use the previously established bounds for standard shadows (Proposition~3 in Ref.~\cite{huang2020predicting})
\begin{align}
    \expp_U[g_\rho(u)]
    \le 
    \expp_U\left[\sum_{\s} 
P_\rho(\s|U) \Big( [\mathcal{M}^{-1}(O^{(1)})](U,\s) \Big)^2 \right]
\le 2^{N_A} ||O^{(1)}_A||_2^2
\end{align}
Here, we used \begin{align*}
\expp_U\left[\sum_{\s}
P_\rho(\s|U) \Big( [\mathcal{M}^{-1}(O^{(1)})](U,\s) \Big)^2  \right] \leq \max_{\sigma \, \text{state}} \expp_U\left[\sum_{\s}
P_\sigma(\s|U) \Big( [\mathcal{M}^{-1}(O^{(1)})](U,\s) \Big)^2\right] = \|O^{(1)}\|_{\text{shadow}}^2 \leq 2^{N_A}  ||O^{(1)}_A||_2^2
\end{align*}
with the shadow norm defined in Ref.~\cite{huang2020predicting} and employed in Eq.\ (S57) to obtain a bound of the shadow norm in terms of the Hilbert-Schmidt norm.

Summarizing and inserting into Eq.~\eqref{eq:def_V},  we find 
    \begin{align}
    \Var(\hat O)
    &\le
    \frac{n^2||O^{(1)}_A||^2}{N_U}
\left(3^{N_A}
\left\|
\rho_A-\sigma_A
\right\|_2^2\;
   +
   \frac{2^{N_A}}{N_M}
\right)
+\mathcal{O}\left(
\frac{1}{N_U^2}
\right).
\label{eq:boundSM}
\end{align}
Note that the term $||O^{(1)}_A||_2^2$ is state-dependent. In particular, for specific states and MCOs with $n>1$, $\textrm{supp}(O^{(1)})$ could be smaller than $A=\textrm{supp}(O)$, i.e.\ $\textrm{supp}(O^{(1)})\subsetneq A$. In this case, we can obtain a tighter bound by  replacing $A$ ($N_A$) in Eq.~\eqref{eq:boundSM} and Eq.~\eqref{eq:boundMT} of the MT,  with $\textrm{supp}(O^{(1)})$ ($|\textrm{supp}(O^{(1)})|$) respectively.
Lastly, we note that for trace moments $\tr(\rho_A^n)$ for which $O=\tau^{(n)}_A\otimes \mathbb{1}^{\otimes n}_{\bar{A}}$, we have, as shown in the next appendix, $O_A^{(1)}=\rho_A^{n-1}$.

\section{Computing $O_A^{(1)}$ for a shift operator $\tau^{(n)}_A$}
Our aim is to calculate the linear operator $O_A^{(1)}$ for $O=\tau^{(n)}_A \otimes \mathbb{1}^{\otimes n
}_{\bar{A}}$ for  $\tau^{(n)}_A$ being the shift operator on $n$ copies of $A$ as defined in the MT. As argued in the statements following Eq.~\eqref{eq:def_Vk}, we can restrict ourselves entirely to the subsystem $A$. We thus drop the subscript $A$ in the remainder of this appendix.

By definition of $O^{(1)}$ [c.f., below Eq.~\eqref{eq:def_Vk}], 
\begin{align}
O^{(1)} &= \tr_{\{2, \dots, n\}} \left( O_{\rm sym} \left[\mathbb{1}_{2^N} \otimes \rho^{\otimes (n-1)} \right]\right)
	= \frac{1}{n!}  \sum_{\pi \in \mathcal{S}_n}^{} \tr_{\{2, \dots, n\}} \left( W_\pi^\dagger   \tau_n W_\pi  \left[\mathbb{1}_{2^N} \otimes \rho^{\otimes (n-1)}  \right] \right).
\end{align} Writing now $\tau_n = \sum_{\s_1,\ldots,\s_n} \ket{\s_n,\s_1,\ldots,\s_{n-1}}\!\bra{\s_1,\ldots,\s_n}$ and $W_\pi = \sum_{\s_1',\ldots,\s_n'} \ket{\s_1',\ldots,\s_n'}\!\bra{\s_{\pi^{-1}(1)}',\ldots,\s_{\pi^{-1}(n)}'}$, we get \begin{align}
W_\pi^\dag\tau_nW_\pi = \sum_{\s_1,\ldots,\s_n} \ket{\s_{\pi^{-1}(1)-1},\ldots,\s_{\pi^{-1}(n)-1}}\!\bra{\s_{\pi^{-1}(1)},\ldots,\s_{\pi^{-1}(n)}}
\end{align}
with $\s_0\equiv\s_n$. Defining $j = \pi^{-1}(1)$, and noting that $\pi^{-1}(2),\ldots,\pi^{-1}(n)$ give all other values $i\neq j$ from 1 to $n$, we then get
\begin{align}
\tr_{\{2, \ldots, n\}} \big( W_\pi^\dag\tau_nW_\pi [\mathbb{1}_{2^N} \otimes \rho^{\otimes (n-1)}] \big) = \sum_{\s_1,\ldots,\s_n} \ket{\s_{j-1}}\!\bra{\s_j} \Big( \prod_{i\neq j} \bra{\s_i}\rho\ket{\s_{i-1}} \Big).
\end{align}
Reordering all the terms (with the index $i$ going down from $j-1$ to 1, and then from $n$ to $j+1$), we get
\begin{align}
& \tr_{\{2, \ldots, n\}} \big( W_\pi^\dag\tau_nW_\pi [\mathbb{1}_{2^N} \otimes \rho^{\otimes (n-1)}] \big) \notag \\
& = \sum_{\s_1,\ldots,\s_n} \ket{\s_{j-1}} \bra{\s_{j-1}}\rho\ket{\s_{j-2}} \bra{\s_{j-2}}\rho\ket{\s_{j-3}} \cdots \bra{\s_1}\rho\ket{\s_0} \bra{\s_n}\rho\ket{\s_{n-1}} \cdots \bra{\s_{j+2}}\rho\ket{\s_{j+1}} \bra{\s_{j+1}}\rho\ket{\s_j} \bra{\s_j} = \rho^{n-1},
\end{align}
where we used the sum rules $\sum_{\s_i} \ket{\s_i}\!\bra{\s_i}=\mathbb{1}_{2^N}$ (recalling that $\s_0\equiv\s_n$). Hence, after (trivially) averaging over $\pi$,
\begin{align}
O^{(1)} = \rho^{n-1}.
\end{align}

\section{Building CRM shadows with a companion experiment}

In the main text, we assumed that we have \textit{a priori} access to a theoretical state $\sigma$. We will now expand our method to accommodate a more state-agnostic scenario where no such theoretical description is available. Instead, we utilize data gathered from a companion experiment realizing a quantum state $\sigma$ to measure an MCO associated with $\rho$. Our only requirement is that the state $\sigma$ represents an approximation of the state $\rho$, which we can then be used to construct CRM shadows for $\rho$. We note that if we could guarantee $\sigma = \rho$, it would likely be more efficient to simply create independent standard shadows with all available data to estimate MCOs. However, the scenario we envision is one where the companion experiment has not precisely implemented $\rho$. Instead, it has the ability to perform RMs on $\sigma$ with a greater number of random unitaries ($N_U' \gg N_U$)  than in the original experiment implementing $\rho$. This could be due to a faster setup or to the ability to aggregate data from multiple prior experiments.

To construct CRM shadows for $\rho$, we proceed as follows.
\textit{(i)} We construct $N_U$ classical shadows $\hat{\rho}^{(r)}$ from the original experiment with $r = 1, \dots, N_U$.
\textit{(ii)} We construct $N_U$ classical shadows $\hat{\sigma}^{(r)}$ from the companion experiment, using the same unitaries $U^{(r)}$ as in the previous step.
\textit{(iii)} We construct $N_U'$ additional, independently sampled, classical shadows $\hat{\sigma}^{(r)}$, $r=N_U+1,\dots,N_U+N_U'$ from the companion experiment. Note that the time ordering is not important here.
From the data gathered in \textit{(i)}, \textit{(ii)} and \textit{(iii)}, respectively, we form three sets of $t=1,\dots,m$ batch shadows,  $\hat{\rho}^{[t]}$ , $\hat{\sigma}^{[t]}$, $(\hat{\sigma}')^{[t]}$, which we combine as
\begin{equation}
    \hat{\rho}_\sigma^{[t]} = \hat{\rho}^{[t]}
    -\hat{\sigma}^{[t]}+(\hat{\sigma}')^{[t]},
    \label{eq:shiftedshadow_companion}
\end{equation}
and which satisfy the desired property  $\expp[\hat{\rho}_\sigma^{[t]}]=\rho$ .
Here, $\expp$ includes the averaging over the $N_U'$ additional unitaries of the companion experiments for the last term in Eq.~\eqref{eq:shiftedshadow_companion}. The finite value of $N_U'$ introduces statistical errors, c.f., numerical examples below. In the limit $N_U'\gg 1$, the CRM shadows of Eq.~\eqref{eq:shiftedshadow_companion} become equivalent to the ones defined in Eq.~\eqref{eq:shiftedshadow}
(because $(\hat{\sigma}')^{[t]}$ converges to $\sigma$), and the variance bound Eq.~\eqref{eq:boundMT} applies.

\begin{figure}
    \centering
    \includegraphics[width=0.4\columnwidth]{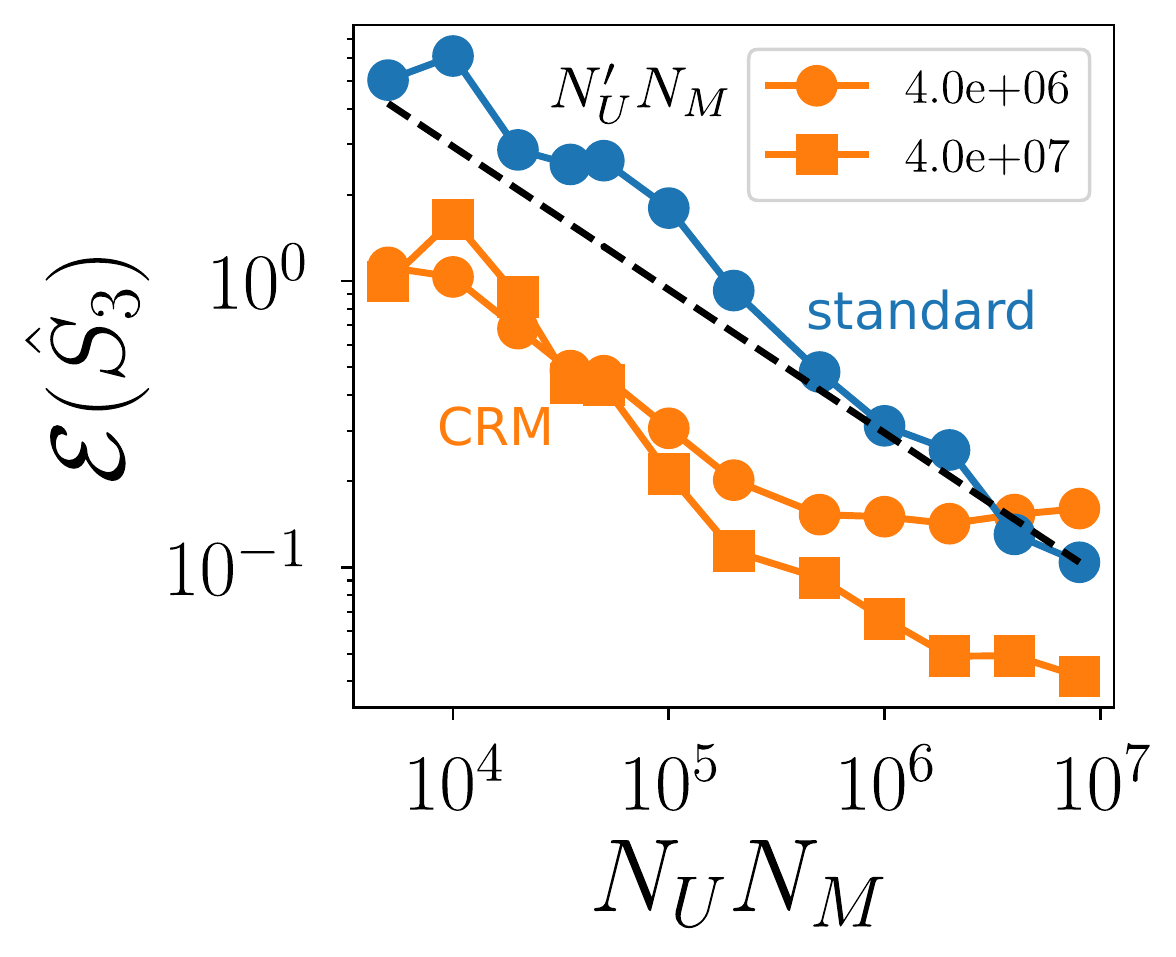}
    \caption{{\it Estimation of the von Neumann entropy in the critical Ising chain, (as in Fig 2 of the MT), but using CRM shadows built from a companion experiment---}
    Relative error $\mathcal{E}(\hat{S}_3)$, where the CRM shadow is formed from a companion experiment using $N_U'$ additional unitaries ($N_M=1000$), as in Eq.~\eqref{eq:shiftedshadow_companion}. Here we use $N_A=6$ ($N=12$). }
    \label{fig:companion}
\end{figure}

In Fig.~\ref{fig:companion}, we show the relative error $\mathcal{E}(\hat{S}_3)$ using CRM shadows built from the companion experiment, considering again the example of the critical Ising chain (as in Fig.~2 of the MT). We use here \mbox{$m=n_{\max}=3$} batches. 
We consider the scenario in which  the ground state $\ket{G'}$ of a Hamiltonian $H'$ implemented in the companion experiment slightly differs from $\ket{G}$ by choosing $H'= H+\sum_i \epsilon_i Z_i$, with $\epsilon_i$ sampled independently in $[0,0.02]$. 
For $N_U'=4000$ (orange circles), we obtain significant error reduction, but we also observe a plateau effect which comes from the finite value of $N_U'$.
When increasing $N_U'$ (orange squares), the plateau's height is reduced, and we obtain excellent CRM estimations compared to standard shadow estimations for all presented values of $N_UN_M$.

\section{Polynomial approximations of the von Neumann entropy via least-square minimization}
\begin{figure}
\includegraphics[width=0.5\columnwidth]{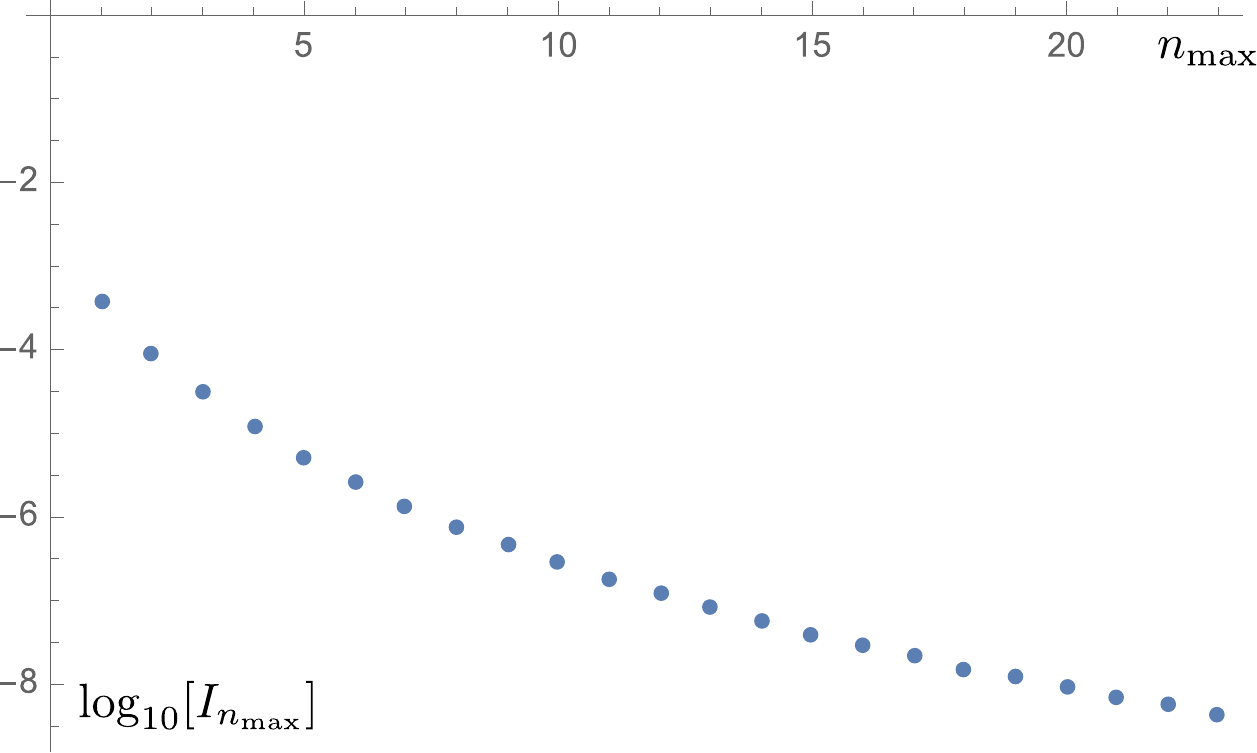}
\caption{Least square error $I_{n_{\max}}$ as a function of $n_{\max}$.}
\label{fig:convergence}
\end{figure}

In this appendix, we explain how to construct the polynomial approximations $S_{n_{\max}}$ introduced in the main text. Our aim is to derive the coefficients $a_n$, $n=1,\dots,n_{\max}$ that minimize the least square error 
\begin{align}
I_{n_{\max}} = \int_{0}^1[f(x)-f_{n_{\max}}(x)]^2 \text{d}x
\end{align}
where $f(x) = - x \log(x)$ and 
$f_{n_{\max}}(x)=\sum_{n=1}^{n_{\max}}a_nx^n$ a polynomial of degree $n_{\max}$.  We find 
\begin{align}
I_{n_{\max}} = -\sum_{n=1}^{n_{\max}} \frac{2a_n}{(2+n)^2}+\sum_{n,n'=1}^{n_{\max}}\frac{a_na_{n'}}{1+n+n'}+\textrm{const},
\end{align}
which we can differentiate
\begin{align}
\frac{\partial I_{n_{\max}}}{\partial a_n}
= -\frac{2}{(2+n)^2}+2\sum_{n'=1}^{n_{\max}} \frac{a_{n'}}{1+n+n'}=0. \label{eq:der}
\end{align}
It follows immediately that $\frac{\partial^2 I_{n_{\max}}}{\partial a_n^2}>0$.
Eq.~\eqref{eq:der} corresponds to a matrix inversion problem $A.c = Y$ with $A_{n,n'}=(1+n+n')^{-1}$ and $Y_n=1/(2+n)^2$.
The inverse of a Cauchy matrix $A_{n,n'}=1/(x_n+y_{n'})$ appears naturally in polynomial approximation problems and is found to be~\cite{knuth1997theart}
\begin{equation}	
A^{-1}_{n,n'}=
\frac{
\prod_{k=1}^n(x_{n'}+y_k)(x_k+y_{n})
}
{
(x_{n'}+y_n)
\prod_{k\neq n'}(x_{n'}-x_k)
\prod_{k\neq n}(y_n-y_k)
}.
\end{equation}
In our case, $x_{n}=n+1$, $y_{n'}=n'$.
We obtain
\begin{align}
a_n
&=
\sum_{n'} A^{-1}_{n,n'}Y_{n'}
=
\sum_{n'}
\left(
\frac{
\prod_{k=1}^n(n'+1+k)(k+1+n)
}
{
(n+n'+1)
\prod_{k\neq n'}(n'-k)
\prod_{k\neq n}(n-k)
}
\right)
\frac{1}{(2+n')^2} .
\end{align}
Having derived explicit expressions for coefficient $a_n$ that determine $f_{n_{\max}}(x)$, we can quantify convergence aspects via the least-square error $I_{n_{\max}}$, which we plot in Fig.~\ref{fig:convergence}.

Once we have built $f_{n_{\max}}$, we can bound the error $|S-S_{n_{\max}}|$ as follows. First, we can find numerically an upper bound $\alpha_{n_{\max}}$ for the function $|f(x)-f_{n_{\max}}(x)|$ in the interval $[0,1]$. For instance, we find $\alpha_{3,4,5}\approx 0.046,0.028,0.019$. Then, we find
\begin{align}
|S-S_{n_{\max}}|&=
\Bigg|\sum_{\lambda\in \mathrm{Spec}(\rho)}
f(\lambda)-f_{n_{\max}}(\lambda)\Bigg|
\le 
\sum_{\lambda\in \mathrm{Spec}(\rho)}
|f(\lambda)-f_{n_{\max}}(\lambda)|
\nonumber \\
&=
\sum_{\lambda\in \mathrm{Spec}(\rho),\lambda\neq 0}
|f(\lambda)-f_{n_{\max}}(\lambda)|
\le
\alpha_{n_{\max}}
\mathrm{rank}(\rho),
\end{align}
where we have used in the second line that $f(0)=f_{n_{\max}}(0)$. 

\end{document}